# Utilizing A Mathematical Model to Estimate Abortion Decline Scenario

Sutiawan, R[1]
[1]Department of Biostatistics and Population Studies, Universitas Indonesia

**Abstract**

Abortion is one of the biggest causes of maternal deaths, accounting for 15% of maternal deaths in Southeast Asia. The increase in and effectiveness of using contraception are still considered to be the effective method to reduce abortion rate. Data pertaining to abortion incidence and effective efforts to reduce abortion rate in Indonesia is limited and difficult to access. Meanwhile such supporting information is necessary to enable the planning and evaluation of abortion control programs. This paper exemplifies the use of a mathematical model to explain an abortion decline scenario. The model employs determinants proposed by Bongaarts, which include average reproductive period, contraceptive prevalence and effectiveness, total fertility rate (TFR), and intended total fertility rate (ITFR), as well as birth and abortion intervals. The data used is from the 1991-2007 Indonesian Demography and Health Survey *(Survei Demografi dan Kesehatan Indonesia/SDKI)*, and the unit of analysis is women who had been married and aged 15-49 years old. Based on the current contraceptive prevalence level in Indonesia at 59-61%, the estimated total abortion rate is 1.9-2.2. Based on the plot of this total abortion rate, an abortion decline scenario can be estimated. At the current TFR level of 2.6, the required contraceptive prevalence is 69% (9% increase) for a decrease of one abortion case per woman. With a delay of one year in the age of the first marriage and a birth interval of three years, it is estimated that the abortion rate will decline from 3.05 to 0.69 case per woman throughout her reproductive period. Based on the assumption of contraceptive prevalence growth at 1-1.4%, it can be estimated that abortion rate will reach nearly 0 between 2018 and 2022.

## Introduction

Not all pregnancies are expected by women. The World Health Organization (WHO) estimated that of 200 million pregnancies every year, around 38% or 75 million of them are unwanted (Berer, 2000). Unwanted pregnancies can happen for a number of reasons, including women or couples not using contraceptives while no longer intending to have children and ineffective use of contraceptives (contraceptive failure) (WHO, 2000).

It is estimated that around two thirds of women who experience an unwanted pregnancy in the world (around 50 million) will resort to induced abortion, and around 40% of them (20 million) are carried out in an unsafe manner by unqualified personnel in places which do not meet medical requirements (WHO, 2000). All around the world, it is estimated around 20 million unsafe abortion cases are performed every year, 95% of which take place in developing countries (WHO, 1994).

Abortion is defined as the expulsion of a fetus resulted from a conception or fertilization before it can survive outside the womb or prior to 20 weeks' gestation or a fetus weighing less than 500 grams (Williams Obstetrics, 1997). Abortion can happen spontaneously or due to an induction. Spontaneous abortion or miscarriage happens naturally without any external induction to terminate the pregnancy (AVCS International, 1998). Induced abortion is abortion that happens as a result of certain induction methods to terminate the pregnancy (AVCS International, 1998) and can be carried out in a safe and unsafe manner. Unsafe abortion is defined as an attempt to terminate a pregnancy carried out by unqualified personnel or using equipment that does not meet medical standards, or both (WHO, 2004).

In developing countries, abortion complications lead to 50,000 – 100,000 maternal deaths every year. WHO estimated that the proportion of maternal deaths due to abortion complications ranges from around 8% in West Asia to 26% in South America, with a world average of around 13% (Population Reports, 1997). In developing countries, the rate could go up to 60% of all maternal deaths (Akin and Ergor, 1998 cited in Widyantoro, 2004). Abortion complications are the biggest cause of deaths among women in the reproductive age (Population Reports, 1997). The risk of death for women who undergo unsafe abortion is 100 times higher than that of safe abortion (Shane, 1997).

There is no definite data on abortion rate in Indonesia. It is estimated there are 750,000 to 1,000,000 abortion cases per year or 18 cases per 100 pregnancies in Indonesia (Hull, Sarwono, and Widyantoro, 1993). The 1997 SDKI data estimated that 12% of all pregnancies ended in abortion (Pradono et al., 2001). A study by Utomo et al. (2001) recorded that there were 37 abortion cases per year per 1,000 women in the reproductive age.

The largest group contributing to unwanted pregnancies that end in unsafe abortion is those that experience contraceptive failure and married women who do not wish to have any more children but do not have access to contraceptives, or known as the group with the unmet need (Wijono, 2009). Contraceptive failure may occur as, the literature suggests, no single contraception method can 100% effectively prevent pregnancy (Guttmacher Institute, 2006; Uddin, et al., 2006). Sterilization surgery has the lowest risk of failure, while more traditional contraception methods such as natural family planning by monitoring the menstrual cycle and withdrawal (coitus interruptus) have higher risks of failure.

Contraceptive failure is a direct cause of unwanted pregnancy, so intervention is needed in order to decrease abortion rate. The use of contraceptives with the lowest risk of failure can be a strategic intervention that the government may consider. Therefore, family planning program campaigns can emphasize on the use of effective contraception methods to reduce abortion rate.

Considering the big impact of abortion on the health and livelihood of women, efforts to bring the number of abortion rate down are imperative. The use of contraceptives as one of the methods to control birth rate can also become an effective intervention to decrease abortion rate. However, it is important to note that not all contraception methods contribute significantly to reducing unwanted pregnancy cases.

Data paucity regarding abortion cases, characteristics, and control through the use of contraception methods is one of the challenges in program evaluation. In developing countries where abortion remains illegal and conservative values remain to prevail in the society, abortion is viewed as a defiant behavior ideologically, socially, and culturally. Within such cultural contexts, it is extremely difficult to directly collect data and information as the society tend to keep sensitive and private matters to themselves. Meanwhile, this information is prerequisite to formulate program planning and evaluation to reduce abortion rate.

**Methods**

Abortion rates estimated include induced abortion and spontaneous abortion (due to natural or health-related causes). This study uses a mathematical model proposed by Bongaarts. Total abortion rate was estimated using a basic regression equation as follows:

$$TAR = \frac{Y_{r(1-e \cdot u)}}{I_A} - \frac{TFR \cdot I_B}{I_A}$$

The total abortion rate (TAR) regression equation includes parameters of direct abortion determinants, namely reproductive period ($Y_r$), contraceptive prevalence (u), contraception effectiveness (e), abortion interval ($I_A$), total fertility rate (TFR), and birth interval ($I_B$).

where:
(1) $Yr = Yr*e*u+TFR*I_B+TAR*I_A$
(2) $TFR=UITFR+ITFR$; ITFR refers to intended TFR
(3) $TAR = Yr(1–e*u)/I_A–TFR*I_B/I_A$
(4) $TAR = p*(Yr*(1–e*u)–ITFR*I_B)/(p*I_A+(1–p)*I_B)$
(5) $p$abortion=$TAR/(TAR+UITFR)$; UITFR refers to unintended TFR
(6) $u_p = e*[(1 - (WTFR*I_B)/Yr)]$

| | |
|---|---|
| ITFR | Average number of children wanted by a woman and pregnancy/birth time as planned |
| TFR | Average number of children per woman throughout her reproductive period |
| WTFR | Average number of children wanted by a woman throughout her reproductive period (including planned and unplanned birth time) |
| Yr | Average period (in years) for each woman aged 15-49 to be pregnant; [35 – (pregnancy+postpartum infecundability+lactation)]*proportion of contraceptive users |
| e | e=1-number of contraceptive failure (pregnancy happens during contraceptive use) |
| u | Contraceptive prevalence (women who use contraceptives) |
| $I_A$ | Time interval in between abortions (Bongaarts recommendation: 14 months) |
| $I_B$ | Time interval in between births (between the last child and the child before) |
| p | Probability of pregnancies that end in abortion |
| $u_p$ | Effective contraceptive prevalence (effective contraception and continuous use of contraceptives) |

Determinants were measured using data from the 1991-2007 Indonesian Demography and Health Survey *(Survei Demografi dan Kesehatan Indonesia/SDKI)*, and the unit of analysis was the number of women who had been married and aged 15-49 years old. The results were then used to estimate the total abortion rate. By modifying the value of certain determinants, the extent to which the changes in the value of the determinants would influence the changes in the total abortion rate could be estimated. This method was used to model the scenario to decrease abortion rate by modifying the value of the determinants.

The projection of total abortion rate decline was plotted based on the contraceptive prevalence growth calculated using an arithmetic growth method:

$$u_t = u_o ( 1 + r (t-o) )$$

whereby r means the average growth of contraceptive prevalence per year. The resulting contraceptive prevalence was used to estimate the total abortion rate using the Bongaarts model.

**Results and Discussion**
Effective contraceptive prevalence was estimated by assuming TAR=0, e=1 (contraceptive effectiveness), and TFR=WTFR. The results of abortion determinants are presented in Table 1.

Table 1. Estimated TFR, TAR, and abortion determinants in several periods according to the survey

| Abortion Determinants | 1991 | 1994 | 1997 | 2002 | 2007 |
|---|---|---|---|---|---|
| Total Fertility Rate (TFR) | 2,8 | 2,7 | 2,61 | 2,55 | 2,602 |
| Total Abortion Rate (TAR) | 4,6 | 3,8 | 2,54 | 2,38 | 1,75 |
| Wanted Total Fertility Rate (WTFR) | 2,7 | 2,51 | 2,46 | 2,4 | 2,46 |
| Intended Total Fertility Rate (ITFR) | 2,3 | 2,22 | 2,15 | 2,19 | 2,13 |
| Abortion Probability ($p$) | 0,9 | 0,9 | 0,85 | 0,83 | 0,8 |
| Reproductive Years ($Y_R$) | 22,7 | 22,6 | 22,5 | 22,3 | 22,1 |
| Average Birth Interval ($I_B$) | 2,3 | 2,4 | 2,5 | 2,7 | 2,9 |
| Contraceptive Prevalence ($u$) | 0,49 | 0,52 | 0,55 | 0,59 | 0,59 |
| Perfective Contraceptive Prev ($u_p$) | 0,77 | 0,71 | 0,71 | 0,7 | 0,68 |
| Contraceptive Effectiveness ($e$) | 0,96 | 0,97 | 0,98 | 0,94 | 0,96 |
| Abortion Interval ($I_A$) | *is assumed: 14 months (Bongaarts)* | | | | |

Source: SDKI 1991-2007

The estimated total abortion rate tended to decrease by an average interval of 1.9-2.2 total abortion cases per woman up to the end of her reproductive period.

Figure 1. The relationship between abortion rate and contraceptive prevalence at several fertility rates

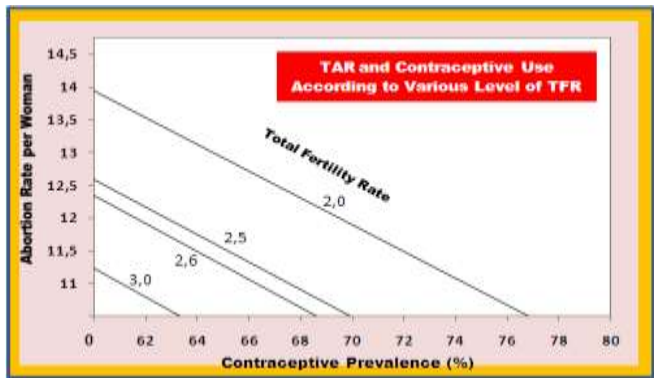

Figure 1 depicts the relationship between contraceptive prevalence and abortion rate at several fertility rates. The relationship was plotted by holding the value of the other determinants constant (e=1, Yr=30, birth interval=2.5, and abortion interval=1.25). As shown in Figure 1, at a total fertility rate of 2.6, abortion rate was close to 0 assuming the contraceptive prevalence was nearly 69%.

Table 2. Estimation model of abortion rate at different levels of abortion determinants

| Determinants | Contraceptive Prevalence ($u$) | | | |
|---|---|---|---|---|
| | 0 | 40 | 60 | 80 |
| **ITFR (Intended Fertility)** | | | | |
| 2 | 4,01 | 2,05 | 1,07 | 0,10 |
| 3 | 3,30 | 1,34 | 0,36 | |
| 4 | 2,58 | 0,63 | | |
| **$Y_R$ (Reproductive Years)** | | | | |
| 17,5 | 2,78 | 1,24 | 0,46 | |
| 20 | 3,40 | 1,63 | 0,74 | |
| 25 | 4,63 | 2,42 | 1,31 | 0,20 |
| **$I_B$ (Birth Interval)** | | | | |
| 2 | 5,63 | 3,12 | 1,87 | 0,61 |
| 2,5 | 4,58 | 2,41 | 1,32 | 0,24 |
| 3 | 3,77 | 1,86 | 0,91 | |
| **$I_A$ (Abortion Interval)** | | | | |
| 1 | 4,08 | 2,04 | 1,02 | |
| 1,25 | 3,84 | 1,92 | 0,96 | |
| 1,5 | 3,62 | 1,81 | 0,91 | |

Table 2 shows the estimation model of abortion rate at different intended fertility rates, reproductive years, birth intervals, and abortion intervals at 0-80% contraceptive prevalence. The tabulation was based on the assumption that abortion probability and contraceptive effectiveness values were held constant at 0.5 and 0.9, respectively.

A negative correlation between fertility rate and abortion rate at each contraceptive prevalence was observed. Close intervals of reproductive years brought negative effects on abortion rates. This means the shorter the reproductive years whereby unwanted pregnancy could happen, the lower the probability of abortion. In contrast, variations in birth and abortion intervals did not result in significant variance of abortion rates. However, in general lengthening the birth and abortion intervals would result in lower abortion rates.

Figure 2. Projection of abortion rate decline assuming a constant contraceptive prevalence growth at 1-1.4% per year

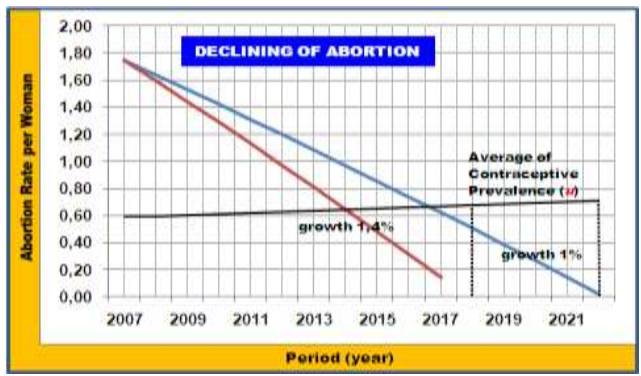

Shortening the reproductive years can be done by increasing the average age of the first marriage and widening the birth intervals. Following this model, an average of one-year increase in the first marriage age and a three-year interval in between births will potentially bring down a total abortion rate from 3.05 to 0.71 per one woman throughout her reproductive period.

The calculation of abortion determinants from the survey in different periods shows that abortion prevalence growth ranged from an average of 1-1.4% annually. Based on this annual growth assumption, the decrease in abortion rate to nearly 0 per woman can be achieved between 2018 and 2022 as shown in Figure 2. Significant increase in contraceptive prevalence growth is needed to reach the lowest possible abortion rate.

Use of contraceptives is considered to be one of the most effective ways to prevent unwanted pregnancy, which can end in abortion. However, not all contraception methods are equally effective in preventing abortion. Abortion rate in the UK increased from 11 cases per 1000 women aged 15-44 years old in 1984 (136,388 cases) to 17.8 cases per 1000 women in 2004 (195,400 cases), when the use of contraceptive pills increased (Glasier A, 2006). A different trend was observed in Russia (Figure 3). Abortion rate decreased by half since the increase in the IUD contraceptive method beginning in 1980 (RAND, 2001). The use of contraceptives in Russia also contributed to the decline in total fertility rate (TFR) to 1.2 per woman aged 35-44 years old, making Russia one of the countries with the lowest TFR. This was also followed by the low abortion rate.

Figure 3. Declining trend of abortion rate in relation to contraceptive prevalence in Russia

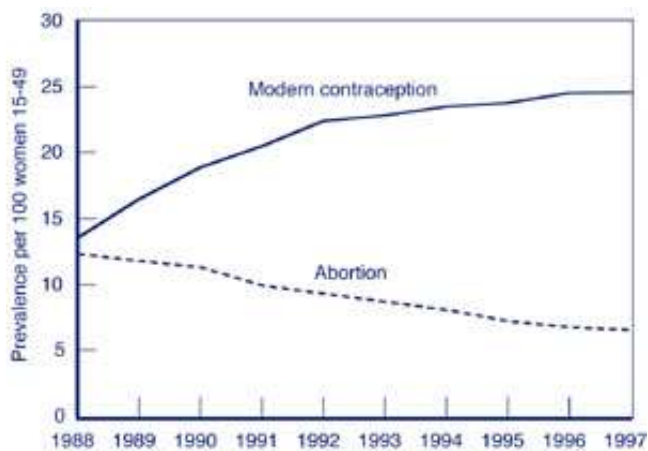

Abortion rate decline in Asian countries generally do not directly follow an increase in contraceptive prevalence. A case in point is Korea, whereby it took about 16 years since the effort to promote the use of contraceptives in 1964 until a significant decline in abortion rate was observed in 1980 (Figure 4).

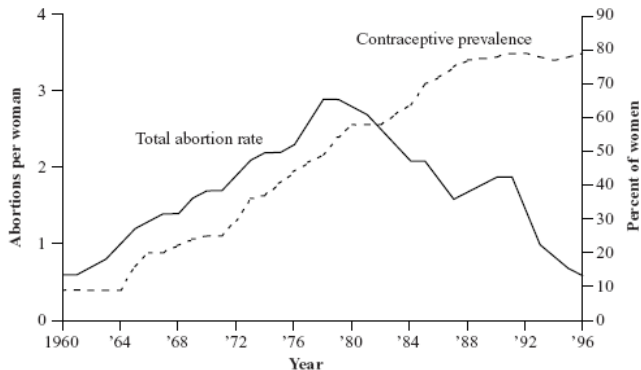

Figure 4. Abortion rate trend in relation to contraceptive prevalence in Korea

Source: United Nation, 1999

**Conclusion and recommendation**

A mathematical model to estimate abortion rates can be applied using the available secondary data from national health survey although this comes with certain limitations. The estimated abortion rate includes total abortion incidence, both induced abortion and spontaneous abortion due to natural causes and health-related reasons. Survey data cannot isolate induced abortion from spontaneous abortion.

Abortion rate was estimated based on seven factors, namely contraceptive prevalence, contraceptive effectiveness, fertility preference, abortion probability, birth interval, abortion interval, and reproductive years. The total abortion rate estimated ranged from 1.9 to 2.2 cases per woman in the reproductive age. Taking a total fertility rate of 2.6 per woman in the reproductive age and annual contraceptive prevalence of 1-1.4%, the estimated total abortion rate was 1.75 per woman in the reproductive age. The average annual contraceptive prevalence growth ranged from 1 to 1.4%. Taking this assumption, the total abortion rate was estimated to be nearly 0 between 2018 and 2022. Based on this model, a scenario to accelerate the total abortion rate decline can be identified using several determinants. To reduce one abortion case per woman, a 9% increase in contraceptive prevalence (u=69%) is required. A one-year delay in the marriage age for women who have never been married and an average of three-year birth interval for married women can potentially decrease the total abortion rate from 3.05 to 0.69 per woman in the reproductive age.

As previous studies suggest that contraceptive use remains the main strategy while its effectiveness to prevent unwanted pregnancy takes considerable time, further studies investigating the extent to which different contraception methods can effectively prevent potential abortion are needed. Such information is necessary to be able to plan and evaluate programs related to effective contraceptive use.

Studies estimating abortion rates using secondary data are bound to have certain limitations, particularly those related to data and information bias. Therefore, studies examining the validity of using secondary data as primary data in studies estimating abortion rates are recommended. Evaluation and studies comparing estimation results with other studies are also needed to assess the measurement tolerance of the estimation results in order to make them useful in program planning and evaluation.

**References:**

1. Berer, M. "Making Abortions Safe: a matter of good public health policy and practice." dalam *Bulletin of WHO 2000, 78(5):580-588*
2. WHO. "*WHO Technical Consultation Safe Abortion: Technical and Policy Guidance for Health Systems Geneva, 18-22 September 2000*". Geneva: World Health Organization, 2000
3. WHO. "*Unsafe abortion: Global and regional estimates of incidence of a*

*mortality due to unsafe abortion with a listing of available country data (Third Edition)"* 2004. dalam www.who.org. di akses tanggal 9 Oktober 2012.
4. Williams Obtetrics 20th edition. Conneticut: Appleton & Lange, 1997: 582.
5. AVSC International. *Penatalaksanaan Klinik Pascaabortus dan Komplikasinya*. Jakarta: AVSC International, 1998.
6. Population Report Series L, Number 10, September 1997. "The Extent of Unsafe Abortion" dalam *http://www.infoforhealth.org/pr/110/110Chap1_2*. di akses tanggal 28 September 2012
7. Akin, Ayse & G. Ergor. *Turkish Expriences on Unwanted Pregnancies and Induced Abortion.* Ankara: Hacettepe University, Medical School, Department of Public Health, 1998.
8. Hull, T.H., S.W. Sarwono and N. Widyantoro. "Induced Abortion in Indonesia". *In Studies Family Planning,* 1993; 24(4): 241-251
9. Affandi, Biran dan S. Sarwono. "Keluarga Berencana dan Aborsi", *Dalam Keluarga Berencana dari Perspektif Perempuan.* Jakarta: YLKI dan The Ford Foundation, 1995.
10. Shane, Barbara. *Family Planning Saves Lives*. Population Reference Bureau (PRB), 1997
11. Utomo, Budi dkk. *Incidence and Socio-Psychological Aspects of Abortion in Indonesia: A Community-Based Survey in 10 Major Cities and 6 Districts, Year 2000*. Jakarta: Center for Health Research University of Indonesia, 2001.
12. Wijono, Wibisono. "Dampak Kesehatan Aborsi Tidak Aman". *Simposium Masalah Aborsi di Indonesia, Jakarta 1 April 2000.*
13. Guttmacher Institute."Facts on Induced Abortion in The United States". May 4, 2006.
14. Bongaarts, John & C.F. Westoff. 'The Potensial Role of Contraception in Reducing Abortion". *Studies in Family Planning Vol.31, No.3, September 2000.*
15. Bongaarts, John. 'Trends in Unwanted Childbearing in the Developing World'. *Working Papers Population Council 1997 No. 98.*
16. Westoff, Charles F. 2008. *A New Approach to Estimating Abortion Rates*. DHS Analytical Studies No. 13. Calverton, Maryland, USA: Macro International Inc.
17. Anna Glasier, James Trussell, 2006. British Medical Journal
18. RAND, 2001. Improvements in Contraception Are Reducing Historically High Abortion Rates in Russia
19. Addor V, F. Narring, and P.A. Michaud. "Abortion trends 1990-1999 in a Swiss Region and Determinants of Abortion recurrence". *Swiss Med Wkly 2003; 133: 219-226.*
20. CBS, NFPCB, MOH, and DHS-Macro Inc. *Demographic and Health Survey (DHS) 2002-2003*. Jakarta: CBS, NFPCB, MOH, and DHS-Macro Inc., September 2002.
21. http://www.populationaction.org/resources/factsheets/factsheet29.htm "Contraceptive Use Helps Reduce the Incidence of Abortion".2005. di akses tanggal 28 September 2012
22. United Nations, 1999. *Levels and Trends of Contraceptive Use as Assessed in 1998*. New York: United Nations, Department for Economic and Social Affairs, Population Division.